\documentclass[aps,pre,reprint]{revtex4-1}
\usepackage{float}
\usepackage{amsmath}
\usepackage{amsfonts}
\usepackage{graphicx}
\usepackage{epstopdf}
\usepackage{epsfig,color}

\def\ee{\mathrm{e}}     
\def\ii{\mathrm{i}}     
\def\sech{\mathop{\mathrm{sech}}}
\def\eps{\varepsilon}
\def\pa{\partial}
\def\Z{\mathbb{Z}}

\begin{document}
\title{Discrete Breathers in a  Mass-in-Mass Chain with Hertzian Local Resonators}

%
%
%
%
%
%

\author{S. P. Wallen \textsuperscript{1}, J. Lee\textsuperscript{2}, D. Mei\textsuperscript{2}, C. Chong\textsuperscript{3}, P. G. Kevrekidis\textsuperscript{2}, and N. Boechler\textsuperscript{1}\\
\textsuperscript{1}Department of Mechanical Engineering, University of Washington, Seattle, Washington 98195, USA\\
\textsuperscript{2}Department of Mathematics and Statistics, University of Massachusetts, Amherst, Massachusetts 01003-4515, USA \\
\textsuperscript{3}Department of Mathematics, Bowdoin College, Brunswick, ME 04011, USA
}

\begin{abstract}
	We report on the existence of discrete breathers in a one-dimensional, mass-in-mass chain with linear intersite coupling and nonlinear Hertzian local resonators, which is motivated by recent studies of the dynamics of microspheres adhered to elastic substrates. After predicting theoretically the existence of the discrete breathers in the continuum and anticontinuum limits of intersite coupling, we use numerical continuation to compute a family of breathers interpolating between the two regimes in a finite chain, where the displacement profiles of the breathers are localized around one or two lattice sites. We then analyze the frequency-amplitude dependence of the breathers by performing numerical continuation on a linear eigenmode (vanishing amplitude) solution of the system near the upper band gap edge. Finally, we use direct numerical integration of the equations of motion to demonstrate the formation and evolution of the identified localized modes, including within settings that may be relevant to future experimental studies.
\end{abstract}

\maketitle

\section{Introduction}
Materials bearing local resonators are known to support unique dynamic phenomena, such as negative, highly anisotropic, and extreme effective properties \cite{MetaBook, Ruzzene}. Systems exhibiting these phenomena are commonly referred to as ``locally resonant metamaterials,'' and are often described using linear dynamical models. One well-known model is a chain of ``mass-in-mass'' unit cells, which consists of a lattice of interconnected lumped masses, each with coupled local resonators \cite{Sun2009}. By incorporating nonlinearity into locally resonant metamaterials, their dynamics become more complex.
For example, metamaterials for both acoustic \cite{KivsharReview, Ruzzene} and electromagnetic \cite{KivsharReview} waves have demonstrated numerous nonlinear phenomena, including tunability, harmonic generation, and the existence of nonlinear localized modes.
	
	A promising means to create nonlinear acoustic metamaterials is provided by granular media, which consist of closely packed systems of particles that interact elastically. Granular media have been shown to support a wide range of nonlinear dynamic phenomena not encountered in conventional materials \cite{NesterenkoBook, Sen08, Kevrekidis2011, GranularCrystalReviewChapter}. In granular materials, the microstructural geometric nonlinearity that stems from the shape of particles in contact (commonly modeled using Hertzian contact mechanics \cite{Hertz}) results in an effective macroscopic nonlinear material response. Previous works on granular media have demonstrated numerous nonlinear effects, including solitary waves, shocks, discrete breathers, tunable band gaps, frequency conversion, and non-reciprocal wave propagation \cite{NesterenkoBook, Sen08, Kevrekidis2011, GranularCrystalReviewChapter}. 
 
Recent theoretical and experimental works have combined the concepts of locally resonant metamaterials with granular media. Relevant contexts include, but are not
limited to, frequency shifting, harmonic generation, and localized band gap modes \cite{Daraio2015PRE}, traveling waves, including ones with non-vanishing tails
 \cite{Stefanov2015,jkprl}, wave interaction \cite{localres_GC2}, and localized and extended modes \cite{localres_GC3}, as well as temporally
periodic breathing states~\cite{Liu2015,Liu2016} (to which we will return in what follows). In each of these examples, the granular media provided a nonlinear intersite coupling, while the local resonators were linear. Less attention has been paid to cases in which granular particles play the role of nonlinear local resonators.

In this work, we consider a one-dimensional, mass-in-mass system with linear intersite coupling and nonlinear local resonators that follow the Hertzian contact model. One motivation for considering this model is its relevance in describing a granular metamaterial consisting of a monolayer of microscale spheres adhered to a substrate, wherein surface localized elastic waves, such as Rayleigh surface acoustic waves (SAWs) and Lamb modes, have been shown to hybridize with the contact resonances of the microspheres in thick \cite{Boechler2013PRL,Hiraiwa2016arXiv} and thin \cite{Khanolkar2015APL} substrates, respectively. Within this context, we imagine the portion of the substrate through which the localized elastic wave is traveling as a linearly coupled chain that is locally coupled to an array of nonlinear resonators representing the microspheres.  

Systems similar to the one-dimensional, linearly coupled chain with nonlinear local resonators considered here have been previously explored. For example, amplitude-dependent band-gaps have been studied in a one-dimensional linear chain with local resonators containing a cubic nonlinearity \cite{Jensen}. Other relevant works have also considered linear chains with nonlinear coupling to a rigid foundation \cite{Vakakis1994}, or a nonlinear local attachment \cite{Vakakis2001}, demonstrating heavily enriched dynamics caused by small nonlinear perturbations.

 The structure of interest in the present work is the discrete breather (DB). Discrete breathers are solutions that are 
periodically oscillating in time and exponentially localized in space~\cite{aubry,Flach1998} that have been studied theoretically and experimentally in many settings, involving a wide array of physical mechanisms \cite{KivsharDBReview,FlachPR2008}. More recently, DBs have been demonstrated in theoretical \cite{Theocharis2010PRE, Vakakis2012} and experimental \cite{Boechler2010PRL, chong14, Vakakis2015} studies of ordered granular chains without local resonators,
and theoretically in the presence of linear local resonators \cite{Liu2015,Liu2016}, as well as in nonlinear, locally resonant magnetic metamaterials \cite{Tsironis,tsironis2a,tsironis2} and systems of electromechanical resonators~\cite{hadi}. 
	
We use our model to describe a locally resonant granular metamaterial for Rayleigh SAWs, consisting of a monolayer of microspheres adhered to a thick elastic substrate. 
The two independent model parameters are fit to an experimental system used in past work \cite{Boechler2013PRL}, so as to provide realistic parameter values to the model wherever possible.
We predict the existence of DBs in the extreme limits of vanishing and strong intersite coupling, and numerically compute a family of DBs connecting the two regimes. 
Furthermore, we examine the frequency-energy dependence of the DBs along the relevant branch of solutions. We study the formation and evolution of the discrete breathers via direct numerical simulations, including within contexts that may be relevant to future experimental studies.
	
\section{Model}

	\subsection{Physical Setup}
	
\begin{figure}[h]
			\centering
			\includegraphics[width=\columnwidth]{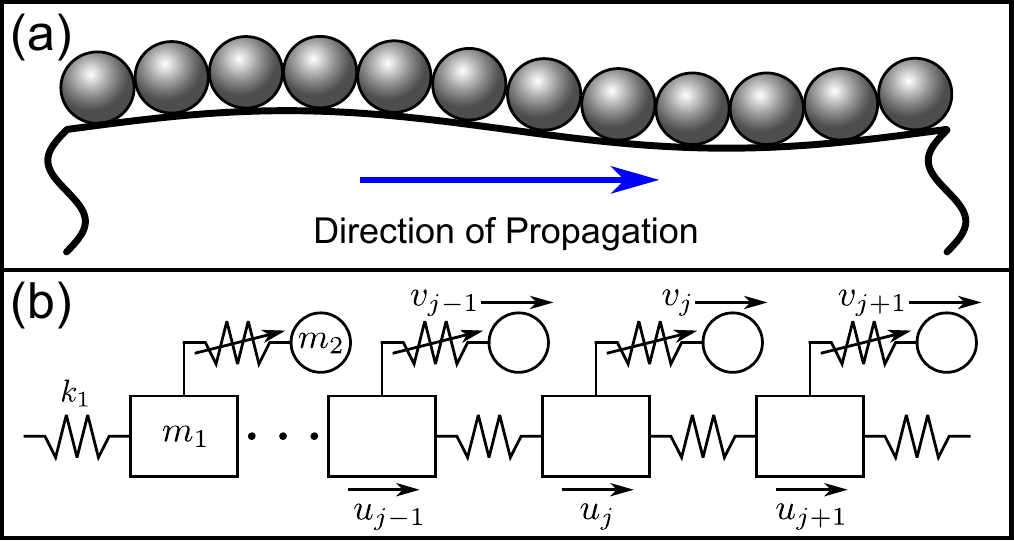}
			\caption{(a) Granular metamaterial composed of a monolayer of spheres on an elastic halfspace. (b) Schematic of the 1D, discrete granular metamaterial model.}
			\label{Fig_MS_Model}
		\end{figure}

Our motivating physical scenario is shown in the schematic of Fig.~ \ref{Fig_MS_Model}(a), which describes sagittally-polarized, plane SAWs traveling along the surface of a thick substrate. Rayleigh SAWs are surface localized elastic waves that travel along a solid surface (represented as an elastic half space in theoretical descriptions), and have both in- and out-of-plane (with respect to the sample surface plane) displacement components \cite{Ewing}. Previous studies on monolayers of microspheres adhered to thick substrates have shown that Rayleigh SAWs in the substrate hybridize with, and excite, microsphere contact resonances having translational out-of-plane \cite{Boechler2013PRL}, and coupled, in-plane, translational and rotational motion \cite{Hiraiwa2016arXiv, Wallen2015PRB}. The hybridization with each of these resonances leads to classic ``avoided crossing'' phenomena \cite{Wigner} characteristic of locally resonant metamaterials and mass-in-mass chains. For the analysis herein, we focus on the avoided crossing with the contact resonance having solely out-of-plane motion \cite{Boechler2013PRL,Hiraiwa2016arXiv, Wallen2015PRB}. Because a plane SAW is confined to the surface of the medium, it can be considered as traveling in one dimension, and as such, we represent the portion of the substrate through which the SAW is traveling as an infinite lattice of lumped masses \(m_1\) connected by springs with linear stiffness \(k_1\). Because the contact-based modes of the microspheres \cite{Cetinkaya2005,Audoin2012,Boechler2013PRL,Hiraiwa2016arXiv} have frequencies much lower than the intrinsic spheroidal vibrational frequencies of the isolated spheres \cite{Sato1962} (e.g. for the microspheres studied in Ref. \cite{Boechler2013PRL}, the out-of-plane contact resonance was measured to be $215$ MHz, while the spheroidal resonance was predicted to be $2.9$ GHz), we model the microspheres as point masses (of mass \(m_2\)) connected to the main chain by nonlinear springs modeling Hertzian contact with a static adhesive load. 
The resulting discrete model of our locally resonant granular metamaterial is shown in Fig.~ \ref{Fig_MS_Model}(b). As can seen in Fig.~ \ref{Fig_MS_Model}(b), the chain elements are both drawn such that their motion is in the horizontal direction. We note that this depiction simply represents the coupling between a substrate (or chain) and a resonator, each having a single degree of freedom with the same, albeit arbitrary, direction of motion. Within the context of the previously described physical scenario, this degree of freedom represents out-of-plane motion of the substrate and the microsphere, as the SAWs propagate along the sample surface indexed by $j$. 
		 
		
	\subsection{1D Discrete Model}
	

In dimensionless form, the associated 
equations of motion of this system read
		\begin{eqnarray}
			  M\ddot{u}_j + K\left(-u_{j+1} + 2u_j - u_{j-1}\right) & \nonumber  \label{eq: EOM1}\\  
			+ \frac{2}{3} \left({[}u_j - v_j + 1]_+^{3/2} - 1 \right) &= 0  \\
			\ddot{v}_j - \frac{2}{3} \left([u_j - v_j + 1]_+^{3/2} - 1 \right) & = 0, \label{eq: EOM2}	
		\end{eqnarray}
	
		\noindent
		where \(u_j\) and \(v_j\) are, respectively, the displacements from equilibrium of the main chain and resonators, \(M=\frac{m_1}{m_2}\),
		 \(K=\frac{k_1}{(3/2) A\sqrt{\delta_0}}\) characterizes
the relative strength of the elastic and Hertzian terms, where the Hertzian coefficient \(A\) depends on the geometry and material properties of the particles in contact \cite{Hertz}, and \(\delta_0\) is the static overlap induced by the adhesive force at equilibrium. The dimensionless time variable \(\tau\) is defined in terms of the physical time \(t\) by \(\tau = \omega_0^{hs} t\), where \(\omega_0^{hs} = \sqrt{(3/2) A\sqrt{\delta_0}/m_2}\) is the resonant frequency of the local oscillator on the elastic halfspace, and the displacements \(u_j\) and \(v_j\) are normalized to \(\delta_0\).  The subscript \([\ \ ]_+\) indicates that the contact force vanishes for resonators that lose contact with the main chain, i.e. when the relative displacement \(v_j - u_j\) exceeds the static overlap. The Hamiltonian (energy) corresponding to Eqs.~\eqref{eq: EOM1} and \eqref{eq: EOM2} is
		\begin{eqnarray}
		  H =&& \sum_{j} \left[ M \frac{\dot{u_j}^2}{2} + \frac{\dot{v_j}^2}{2} + \frac{K}{2}(u_{j+1}^2 - 2u_{j+1}u_j + u_j^2) + \right. \nonumber \\ 
		  && \left. \frac{2}{3}\left( \frac{2}{5}[u_j - v_j + 1]^{5/2}_+ - (u_j-v_j) \right) - \frac{4}{15} \right] .  \label{ham}
		\end{eqnarray}
%
%
		
%
		Upon linearization, this system is identical to the one-dimensional mass-in-mass chain discussed in \cite{Sun2009}. Its dispersion relation is given by
			\begin{eqnarray}
				M\left(\frac{\omega}{\omega_0^{hs}}\right)^4 && - [2K(1-\cos{(kD)}) + M+1]\left(\frac{\omega}{\omega_0^{hs}}\right)^2 \nonumber \\
				&& + 2K(1-\cos{(kD)}) = 0, \label{eq: DR}
			\end{eqnarray}
		
		\noindent where $ k $ is the Bloch wave number, $ \omega $ is the angular frequency, and $ D $ is the unit cell width, taken from the physical system as the width of the granular particles. 
%

	\subsection{Parameter Fitting} \label{sec:param_fit}
		
\begin{figure}[t]
			\centering
			\includegraphics[width=\columnwidth]{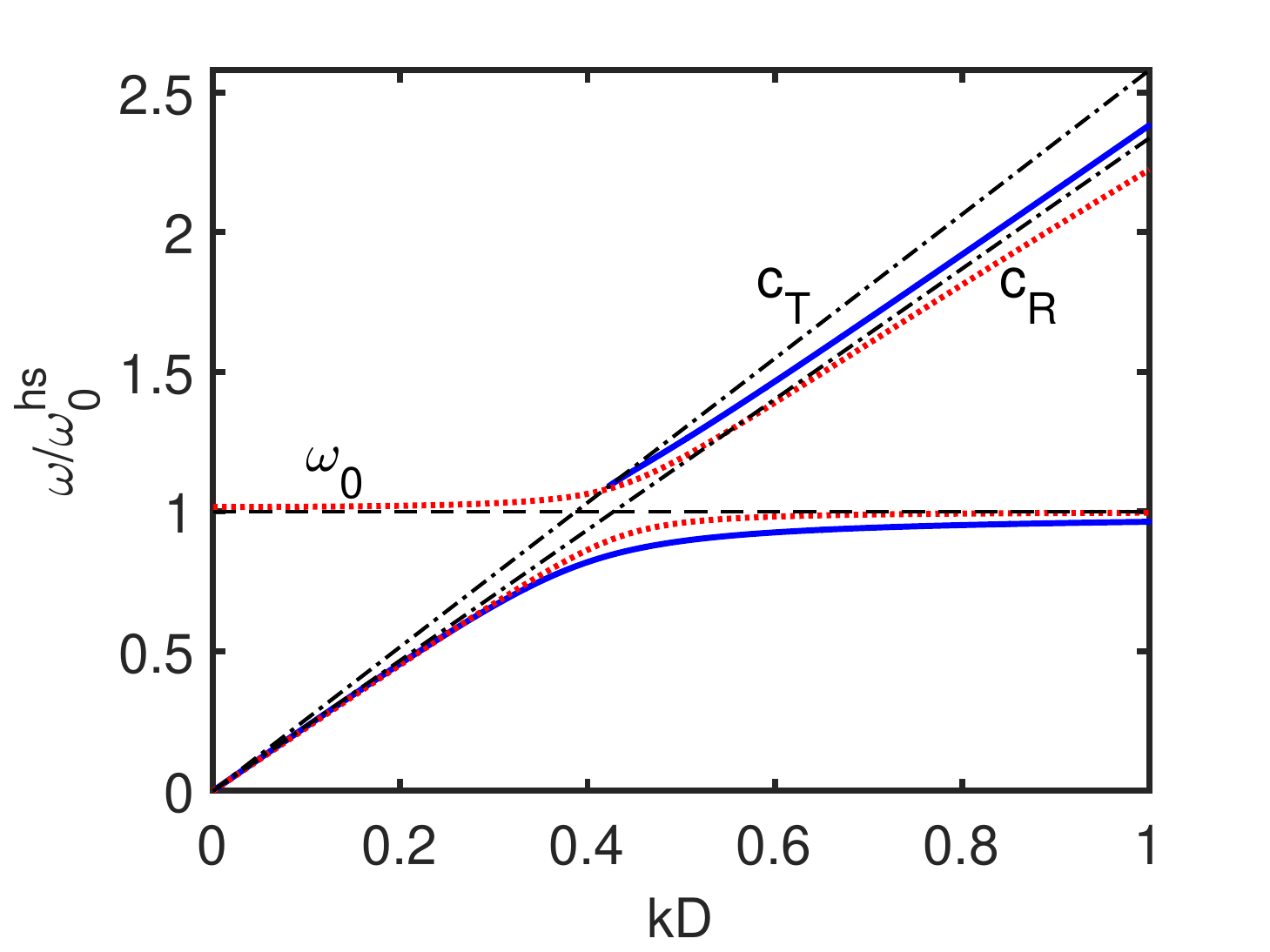}
			\caption{Blue solid and red dotted curves denote the dispersion relations for the continuous and discrete granular metamaterial models, respectively. Black dash-dotted lines have slopes equal to the transverse ($ c_T $) and Rayleigh ($ c_R $) wave sound speeds of the substrate. The black dashed line denotes the linear natural frequency of the Hertzian local resonators. The discrete model uses the fitted parameters $ K = 30$ and $ M = 160$.}
			\label{Fig_DR_Fitted}
		\end{figure}

The resulting model, as can be inferred from 
Eqs.~(\ref{eq: EOM1})-(\ref{eq: EOM2}), possesses two effective lumped
parameters, namely $M$ and $K$.
We now attempt to fit these discrete model parameters to describe the granular metamaterial of Ref. \cite{Boechler2013PRL}, using the material and geometric properties specified therein. We are intending for the model 
to provide an adequate representation of dynamical evolution for 
wavelengths significantly larger than the sphere diameter, such that the dispersion relations of the continuous and discrete models are in close agreement, and effects found at the Brillouin zone boundary \cite{BrillouinBook} are avoided \cite{Sun2009}. The dispersion relations for a continuous substrate (from Eq. (2) of Ref. \cite{Boechler2013PRL}) and the discrete model of Eqs. \eqref{eq: EOM1} and \eqref{eq: EOM2} are superimposed in Fig.~ \ref{Fig_DR_Fitted}. The values of $M$ and $K$ used to plot the curves for the discrete model shown in Fig.~ \ref{Fig_DR_Fitted} are chosen to match the dispersion curves in the small wavenumber regime. More specifically, we first choose the ratio $ K/M $ such that the long-wavelength sound speed of the discrete lattice, given by $ D\sqrt{K/M} $ \cite{BrillouinBook}, matches the speed of Rayleigh waves in the substrate for the model from Ref. \cite{Boechler2013PRL}. This can be seen graphically in Fig.~ \ref{Fig_DR_Fitted}, as the lower branches of the two dispersion relations have equal slopes at the origin. Second, making use of the analytical expression for the dispersion relation of the continuous system in Ref. \cite{Boechler2013PRL}, we select $ K $, such that the dispersion relations coincide at the intersection with the line of slope $ c_T $, where $ c_T $ is the transverse sound speed of the substrate material. 
Using these two criteria, we find the approximate fitted values $ M = 30 $ and $ K = 160 $.  

The physical significance of these parameter values, $M$ and $K$, is as follows. For large mass ratios ($ M >> 1 $), waves in the main chain (corresponding to Rayleigh SAWs in the substrate) are only perturbed at frequencies very close to the local resonance; this is confirmed in Fig.~ \ref{Fig_DR_Fitted}, by the relatively narrow band gap encompassing $ \omega/\omega_0^{hs} = 1 $. The large stiffness ratio ($ K >> 1 $) indicates strong coupling between lattice sites, compared to the coupling between the main chain and the resonators. Intuitively, parameters much greater than unity are indeed expected for this system. This is because the spheres are much smaller and less massive than the region of the substrate beneath them that is influenced by Rayleigh waves, whose displacements decay exponentially from the surface with a characteristic decay length on the order of one wavelength \cite{Ewing}. Similarly, the effective stiffness for the region of bulk material of the substrate influenced by the Rayleigh wave can be thought to have a significantly greater effective stiffness than the relatively soft microsphere-substrate Hertzian contact. While the fitted constants depend on material and geometric properties, a simple estimate can be used to show that $ M $ and $ K $ are generally larger than unity when considering long waves in realistic materials, as described in Appendices A and B.

Both of the dispersion relations shown in Fig.~ \ref{Fig_DR_Fitted} are split, as a result of the hybridization with the local resonance, into two branches: the lower  (``acoustic'') branch, in which the vertical motions of the substrate surface and the spheres are in phase, and the upper (``optical'') branch, in which the motions are out of phase. The two branches of the discrete model are separated by a band gap of width $ \Delta\omega $, given by
\begin{eqnarray}
	\Delta \omega && = \sqrt{\frac{1+M}{M}}\\ \nonumber
	 &&- \sqrt{\frac{4K+M+1 - \sqrt{(4K+M+1)^2-16KM}}{2M}}.
\end{eqnarray}
\noindent
 For the parameters values $M$ and $K$ used herein, this band gap results in an upper cutoff frequency of the acoustic branch $0.08$\% below $\omega_0^{hs}$ and a lower cutoff frequency of the optical branch $1.65$\% above $\omega_0^{hs}$. In the continuous model, for phase velocities greater than the sound speed of transverse bulk waves ($ c_T $) in the substrate, the optical branch terminates; this is because the modes above that phase velocity are so-called ``leaky'' modes. Such leaky modes have complex frequency, which represents the radiation of energy into the bulk \cite{Ewing}. Despite this, the one-dimensional, discrete model used in this work was chosen over continuous and/or higher-dimensional models because it captures many of the important features of the dispersion relation (linear dispersion at long wavelengths and a band gap created by a local resonance), and facilitates theoretical prediction of discrete breathers, and their numerical computation.

%

\section{Theory}

	\subsection{Anticontinuum Limit}
	
	We start our analysis by considering the so-called anti-continuum (AC) limit of vanishing coupling. This approach, pioneered by MacKay and Aubry in \cite{macaub}, is based on the limit $K\rightarrow 0$, corresponding to uncoupled oscillators. While this limit is of limited physical relevance for
our considerations herein, it is a particularly useful mathematical
tool as a starting point for considering different breather-type 
configurations.
This enables a natural starting point to seed continuation algorithms, which
are then continued in the parameter $K$ in order to 
identify solutions for different values of the relevant
parameter. In the AC limit our system  has the form
		\begin{eqnarray}\label{Eqn:Motion1}
			M\ddot{u}_j &= -\frac{2}{3}\left([u_j - v_j + 1]_+^{3/2} - 1\right) \\
			\ddot{v}_j & = \frac{2}{3}\left([u_j - v_j +1]_+^{3/2}-1\right)\label{Eqn:Motion2}.
		\end{eqnarray}
We obtain a single oscillator by defining $z = u_j - v_j$, where
		\begin{eqnarray}\label{Eqn:Lin}
			\ddot{z} = - \frac{2}{3} \omega_0^2\left([z+1]_+^{3/2} - 1\right), \qquad \omega_0^2 = \frac{1+M}{M}.
		\end{eqnarray}
In addition to the trivial solution $z = 0$, there are non-trivial solutions of Eq.~\eqref{Eqn:Lin} which are the level curves of the energy $E(z,\dot{z}) = \frac 1 2 \dot{z}^2 + \frac{4(1+M)}{15M}\left([1+z]_+^{5/2} - \frac{5}{2}z\right)$.
To construct a solution along the infinite lattice, each node is given as $z=0$ or the periodic function (say with frequency $\omega_b$) satisfying \eqref{Eqn:Lin}. In this paper, we consider the simplest such configuration,
 namely the one consisting of zeros at every node with the exception of one (see Fig.~\ref{NLSbreather}(a)). Due to Ref. \cite{macaub} we know this solution will persist for nonzero $K$ as long as the so-called non-resonance condition
$ \omega_b \neq n \omega_0^*, n\in \Z $
is satisfied, where $\omega_0^* \in \{0,\omega_0\}$ is the frequency of solutions of the linearized equations at $K=0$.  While any such value of $\omega_b$ will yield a persistent breather solution, we
chose $0<\omega_b < \omega_0$. Note that the two branches of the dispersion curves given by Eq. \eqref{eq: DR} bifurcate
from $0$ and $\omega_0$ as the coupling $K$ becomes nonzero.  Thus, by choosing $0<\omega_b < \omega_0$ we are able to construct a band-gap breather.
The numerical continuation (see Sec.~\ref{SectionCont}) suggests that the solution constructed in the AC limit persists to the opposite limit $K\rightarrow \infty$.  In this limit, 
other analytical techniques are available for the analysis of the solutions, which we explore next.

	\subsection{Continuum Limit and NLS Approximation}
\begin{figure}[t]
			\centering
			\includegraphics[width = \columnwidth]{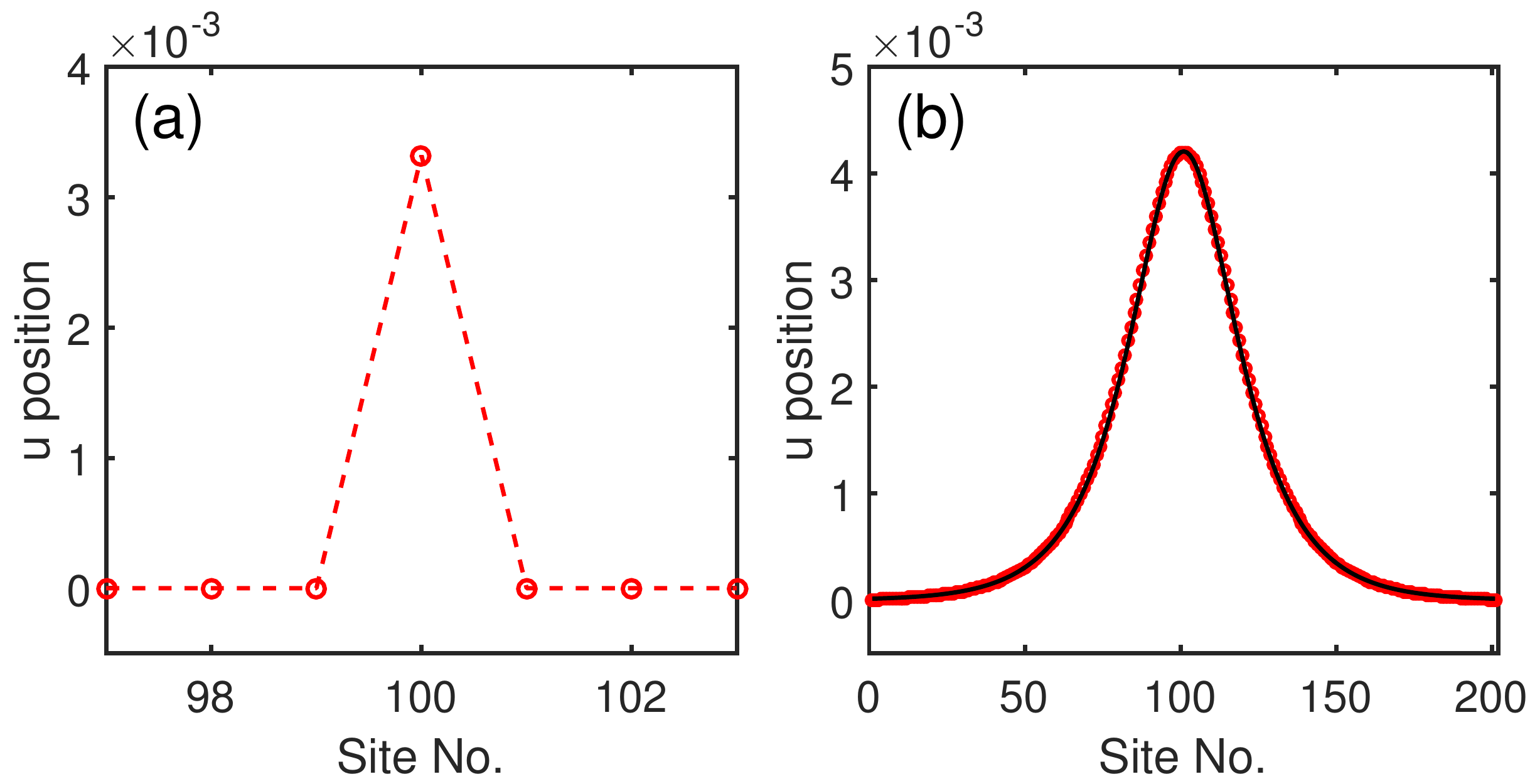}
			\caption{(a) Single site solution in the AC limit with $\omega_a = 1.01\omega_0^{hs}$. (b) Breather profiles of frequency $ \omega_a = 1.01 \omega_0^{hs} $ from NLS ansatz (black solid curve) and numerical solution (red circle markers) with $K=160$. Here, we use $M=30$.}
			\label{NLSbreather}
		\end{figure}

		For the purposes of the analysis, we consider small amplitude solutions (i.e. $|u_j - v_j| \ll 1 $). Thus, it is reasonable 
to expand the nonlinearity in a Taylor series
		$$\left[ 1+x \right]^{3/2} = 1 + \frac{3}{2} x + \frac{3}{8} x^2 - \frac{3}{48} x^3 + h.o.t.$$
     	In addition, if we formally consider $K = 1/D^2 $ where $D$ is the lattice spacing and we let $D\rightarrow 0$, then Eqs. \eqref{eq: EOM1} and \eqref{eq: EOM2} become
     	\begin{eqnarray}
     	M \pa _{tt} u - \partial_{xx}u & \nonumber \\ 
     	+ (u -v) +\frac{1}{4}(u -v)^2 -\frac{1}{24}(u -v)^3  &=& 0 \label{kg1} \\
     	\pa_{tt}v-  (u -v) -\frac{1}{4}(u -v)^2 +\frac{1}{24}(u -v)^3 &=& 0. \label{kg2}
     	\end{eqnarray}
	We approximate solutions of the above set of equations with the ansatz,
     	\[u^{an} = \varepsilon A(X,T)E(x,t) + c.c. + h.o.t. ,   \]
      where $X=\eps (x-ct),\ T = \eps^2t $,  $A=A(X,T)$ and $E=E(x,t)=\ee^{\ii (kx+\omega/\omega_0^{hs} t)}$ with $\eps$ being some small, positive parameter. 
      Substitution of the ansatz into Eqs.~\eqref{kg1} and \eqref{kg2} and equating the various orders of $\epsilon$ yields a hierarchy of solvability
      conditions. The particular choice of ansatz is well known to yield a Nonlinear Schr\"odinger (NLS) equation for the envelope function $A(X,T)$ in the theory of nonlinear waves \cite{ablowitz}. For our
	system, in order to derive the NLS equation we need several higher order terms:
		\begin{eqnarray}
		u^{an} = \eps AE + \eps^2a_2A_1E^2 +\eps^3 a_4A_3E^3 +\eps^3a_6 A_5 &+\nonumber \\
		\eps^3 a_8 A_7E^2 +\eps^3a_{10}A_9E + c.c.&  \label{longanz1}\\
		v^{an} = \eps a_1 AE + \eps^2a_3 A_2E^2 +\eps^3 a_5A_4E^3 +\eps^2a_7 A_6&+ \nonumber \\
		\eps^3 a_9 A_8E^2 +\eps^3a_{11}A_{10}E + c.c. +\eps^2a_{12}A\bar{A},& \label{longanz2}
		\end{eqnarray}  
		where each $A_i = A_i(X,T)$ and the $a_i$ are real or complex coefficients.
 		In particular, we will use the following,
 		\begin{eqnarray}
 		A_1=A^2=A_2 ,\  A_3 =A^3= A_4,\ A_5 =\bar{A}	 	\pa_XA, \nonumber \\
		A_6 =\pa_XA, \  A_7 = A\pa_XA= A_8 ,\ A_9 = \pa_T A, \ A_{10} =\pa_X^2A. \nonumber 
	 	\end{eqnarray}
		These relations are obtained through the solvability conditions, which can be found  
		 in Appendix~\ref{NLS_append}.
		We highlight here that at
		$\mathcal{O}(\eps E)$ the solvability condition is the dispersion relation,
		\begin{equation}
		  M(\omega/\omega_0^{hs})^4 - \left[ k^2+M+1\right](\omega/\omega_0^{hs})^2 + k^2=0, \label{contdisp}
		\end{equation}
		where $k$ is the Bloch wave number and $\omega$ is the angular frequency. The connection between this dispersion relation
		and the one in Eq.~\eqref{eq: DR} can be seen by Taylor expanding the cosine terms of Eq.~\eqref{eq: DR}.
          The Nonlinear Schr\"odinger (NLS) equation appears at $\mathcal{O}(\eps^3 E)$, 
		\begin{eqnarray} \nonumber
		\left((M\omega^2-k^2-1)a_{10}-2iM\omega\right)   \pa_TA 		&= \\
		\left( Mc^2-a_{11}-1\right)\pa_X^2A  &+ \nonumber\\   
		\left(\frac{(1-a_1)}{2}(a_2 -a_3-a_{12}) +\frac{(1-a_1)^3}{8}\right)|A|^2& A.   \label{NLS}
		\end{eqnarray}
		Closed form analytical solutions of the NLS equation \eqref{NLS} can be found via the inverse scattering transform \cite{ablowitz}. One well known solution
		is the so-called bright soliton, and is given by
 		\begin{equation}\label{eq:analytic}
 		A(X,T) = \sqrt{\gamma}  \alpha \sech\left( \sqrt{\gamma}  \beta X \right) \ee^{- i\gamma T},
 		\end{equation}
     	      where $\alpha$ and $\beta$ are $\epsilon$ independent coefficients that depend on the coefficients of the NLS equation \eqref{NLS} (see Appendix~\ref{NLS_append}) and $\gamma>0$ is an arbitrary parameter.
Such solutions arise when the coefficient of the dispersion term
and that of the nonlinearity have the same sign, which is the case for the parameter values chosen here.
%
%
	      Thus, at first order, we have the following approximation 
	      $$ u(x,t) \approx \epsilon \sqrt{\gamma} \alpha \sech\left(\beta \sqrt{\gamma} X \right) \ee^{- i \gamma T}\ee^{\ii (kx+\omega t)} + c.c., $$
	      which is traveling plane wave that is modulated by a small amplitude, long wavelength and slowly varying localized function. For $k=0$, this approximation reduces to
	      \begin{equation}
	       u(x,t) \approx \epsilon \sqrt{\gamma} \alpha \sech\left(\beta  \epsilon \sqrt{\gamma} x \right) \ee^{ i(\omega_0- \gamma \epsilon^2) t} + c.c., \label{NLSap}
	       \end{equation}
	      which represents a standing breather with frequency $\omega_b = \omega_0 - \gamma \epsilon^2$, see Fig.~\ref{NLSbreather}(b). Here $\omega_0^2 = (M+1)/M$ represents the lower cutoff of the optical
	      branch of the dispersion (see Eq.~\eqref{eq: DR} with $k=0$). Since $\epsilon$ is a small parameter, the breather frequency $\omega_b$ is near 
the lower cutoff of the optical branch, but within the gap. Note the amplitude and width of the breather are both $\mathcal{O}( \epsilon \sqrt{\gamma} )$. Hence smaller amplitude and wider breathers are found closer
to the optical branch band edge. Recalling the results of the previous section on the AC limit, we were able to construct a breather solution with frequency 
$\omega_b<\omega_0$. If we define
	      $\gamma \epsilon^2 = \omega_0 - \omega_b$, then our NLS approximation \eqref{NLSap} (which is valid for large $K$) will have, to first order, the same frequency, as the 
	      AC limit breather (which is relevant for small $K$). In the next section, we will perform parametric continuation in $K$ in order to connect the approximations in the two opposing limits $K=0$ and
	      $K \rightarrow \infty$.

\section{Numerical Investigation of Breathers \label{SectionCont}}

\subsection{Continuation in Intersite Coupling Stiffness, \(K\)}
	The analysis above allows us to describe breather solutions of Eqs.~\eqref{eq: EOM1} and \eqref{eq: EOM2} in the limits $K\rightarrow0$ and $K\rightarrow \infty$. In order to connect these two pictures, 
	we identify periodic orbits and explore their parametric continuation~\cite{kuznetsov}.
 Our seed solution (initial guess) will be the AC limit solution with a single excited site, (see Fig.~\ref{NLSbreather}(a)). For the example considered in this section, 
	we use the fitted value $M=30$ from section \ref{sec:param_fit}, and we choose the breather frequency $\omega_b = 1.01\omega_0^{hs}$. Note that in this case $\omega_0 = \sqrt{1 + 1/30}>1.01$. For a fixed breather period $T = 2\pi/\omega_b$,
	we use the fact that $u_j(0) = u_j(T)$ and $v_j(0) = v_j(T)$ to construct the Poincar\'{e} map 
	%
  	 %
  	 \begin{eqnarray}\label{Eqn:Poin}
  	 	P([u_0,v_0];T) &=
          		\begin{bmatrix}
           		\textbf{u}(0;u_0,v_0) \\           
           		\textbf{v}(0;u_0,v_0)
          		\end{bmatrix} -
          		\begin{bmatrix}
          			\textbf{u}(T; u_0,v_0) \\
          			\textbf{v}(T;u_0,v_0)
          		\end{bmatrix}
  	 \end{eqnarray}
	where $\textbf{u}(t;u_0,v_0) = \{u_n(t)\}_{n\in[0,N]}$ and  $\textbf{v}(t;u_0,v_0) = \{v_n(t)\}_{n\in[0,N]}$ is the solution
 to Eqs. \eqref{eq: EOM1} and \eqref{eq: EOM2} with initial condition $u(0) = u_0$ and $v(0) = v_0$. Therefore, a periodic solution with period $T$ of Eqs. \eqref{eq: EOM1} and \eqref{eq: EOM2} will be a root to (\ref{Eqn:Poin}). A Newton-Raphson 
 algorithm is used to approximate the roots of $P$ \cite{Flach1998}. The Jacobian is $J = I - V(T)$, where $V(T)$ is the monodromy matrix.
 The eigenvalues of $V(T)$ are the Floquet multipliers of the periodic solution. The breather is considered (spectrally) stable if all Floquet multipliers lie on the unit circle. Since the system is Hamiltonian, any Floquet multiplier lying off the unit
 circle signals instability \cite{Flach1998}.
 %
%
%
%
  	%
  	 %

\begin{figure}[h]
				\centering
				\includegraphics[width =\linewidth]{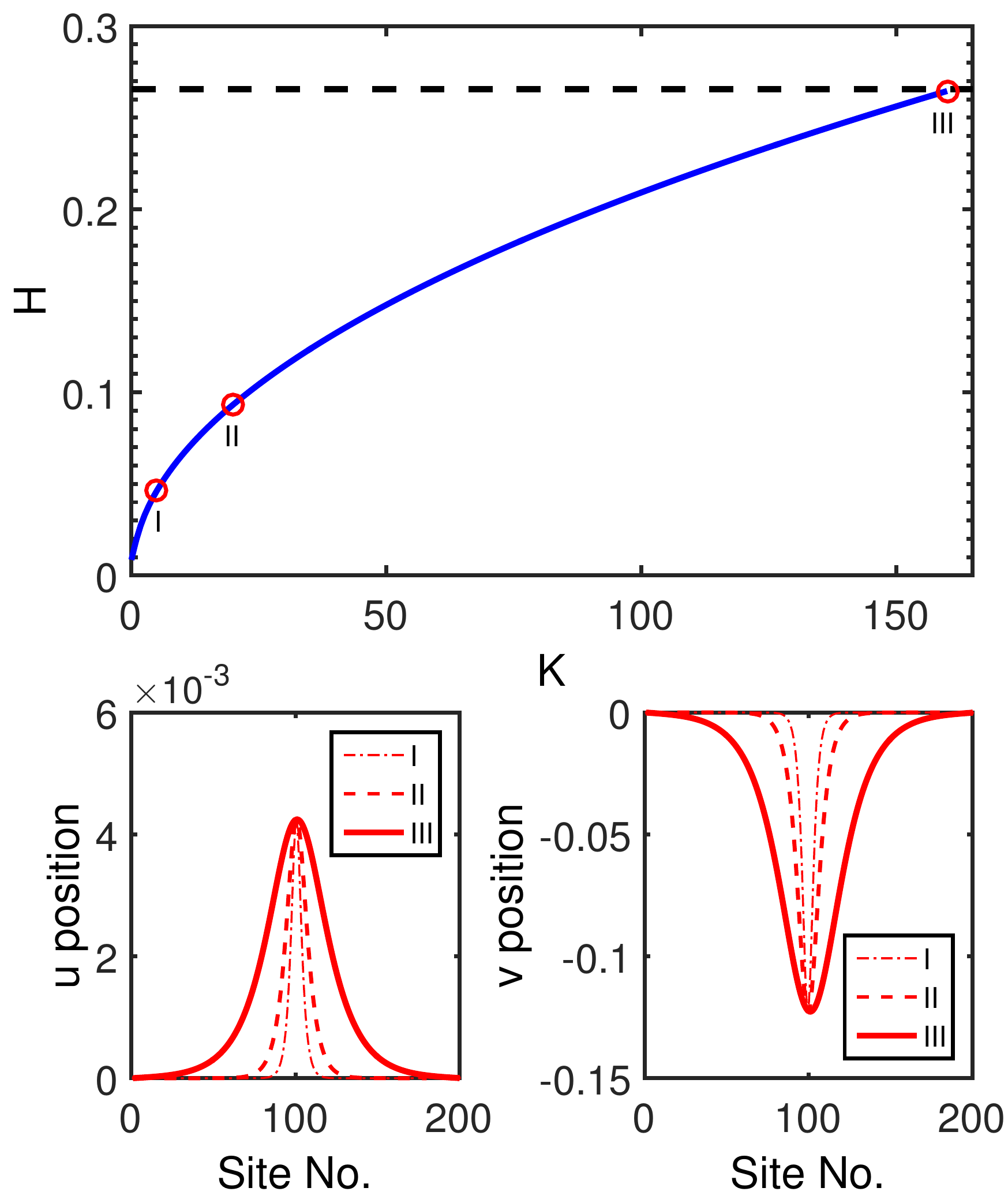}
				\caption{Top panel is the continuation diagram (Hamiltonian energy versus coupling parameter $K$), where the breather frequency and mass ratio are fixed as $\omega_b = 1.01\omega_0^{hs}$ and $M=30$. The dashed black line is the Hamiltonian of the NLS approximation of Eq. \eqref{NLSap}, given by Eq. \eqref{ham}, with K=160. The bottom left and right panels are the main chain displacements  $u_j$ and the local resonator displacements $v_j$, respectively, taken at particular values of parameter $K$: for I, $K= 5$; for II, $K=20$; and for III, $K=160$.}
\label{Breathers_contK}
		\end{figure}

	 The solution for a frequency ($\omega = 1.01\omega_0^{hs}$) is found for $K$ close to $0$ (here $K=0.01$). Parameter continuation in $K$ is performed and is plotted against the Hamiltonian energy of Eq. \eqref{ham}, as is shown in Fig.~ \ref{Breathers_contK}. In this way, we are
	 able to trace out a branch of solutions that emanates from the AC limit and approaches the continuum limit solution, which is well described by the NLS approximation of Eq. \eqref{NLSap} (see black dashed-line of Fig.~ \ref{Breathers_contK}).
	 Here, we terminate the continuation at $K=160$, which corresponds to the stiffness parameter extracted by fitting the discrete model to the locally resonant half space model from Ref. \cite{Boechler2013PRL} (see Sec.~\ref{sec:param_fit}).
The numerical computations were performed on a system of 201 unit cells. This solution was spectrally stable for all values of $K$ considered.

%
%

	\subsection{Continuation in Frequency, \(\omega\) \label{SectionContFreq}} 
		The continuation in $K$ of the previous section was terminated at the parameter value  $ K = 160 $  (note $ M = 30 $ and $\omega_b = 1.01 \omega_0^{hs}$). We now
		fix $M=30$ and $K=160$ and vary the breather frequency
		$\omega_b$ using a pseudo-arclength continuation procedure \cite{NNMs1, NNMs2}. 
		%
		This  allows us to visualize the energy-frequency dependence of the breathers, as may be studied in an experiment with varied excitation amplitude. 
		
		As a seed for the continuation, we use the eigenmode of the linearized system nearest the lowest optical band edge, which is a time-periodic solution of the full nonlinear equations of motion under conditions of vanishing amplitude. We have included a check in our computations to detect a loss of contact between the main chain and resonators, and continue the solution branch until this point.
		
\begin{figure}[h]
			\centering
			\includegraphics[width=\columnwidth]{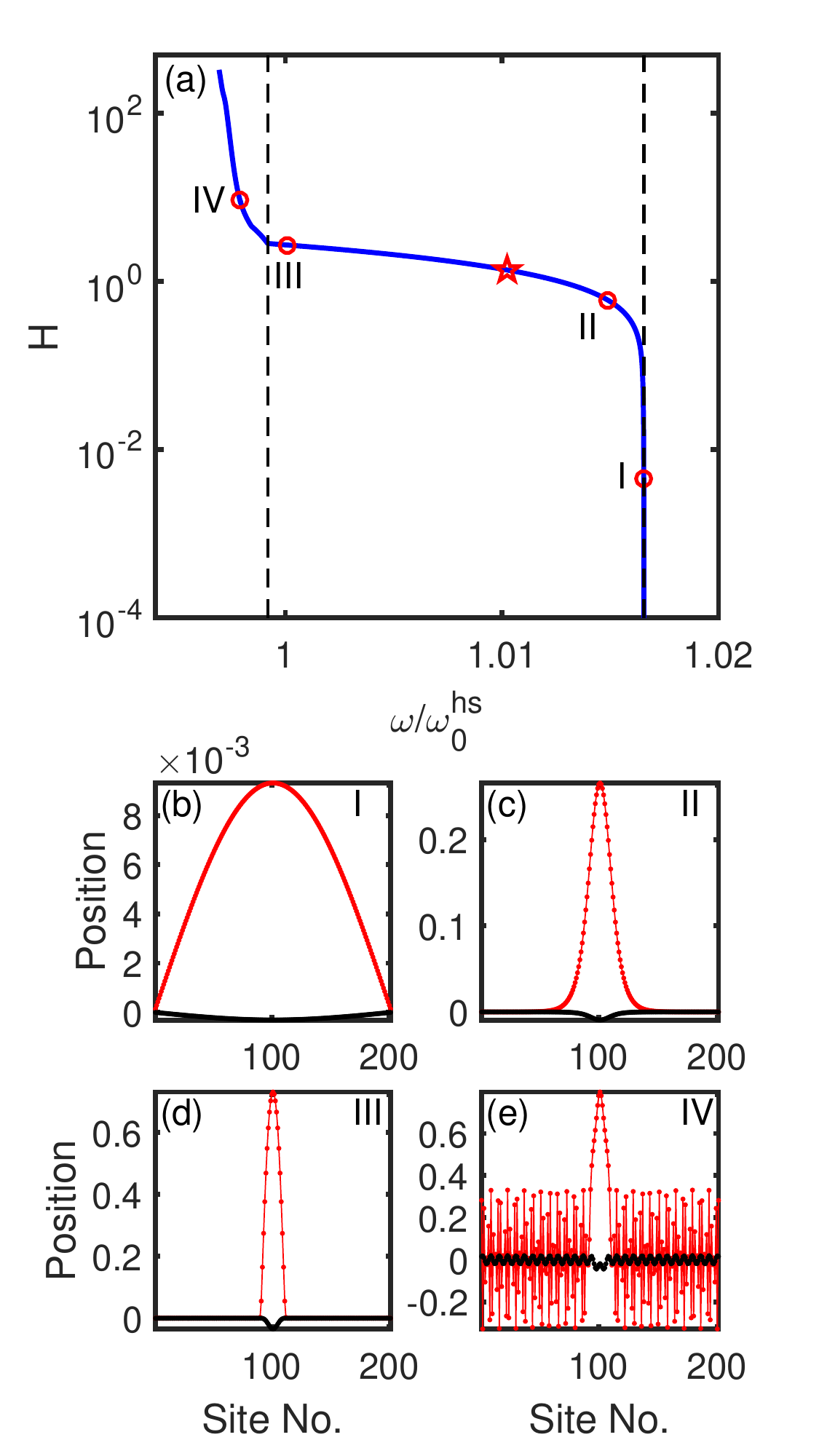}
			\caption{(a) Hamiltonian energy-frequency plot of the family of breather solutions bifurcating from the lowest eigenmode of the optical band. Black dashed lines indicate the edges of the linear phonon band gap. Red star corresponds to the breather shown in Fig. ~\ref{NLSbreather}(b). (b - e) Breather displacement profiles corresponding to the points labeled $ I $ - $ IV $, respectively, in (a).  
		  The main chain displacements $ u_j $ are shown as black points, and those of the local resonators $ v_j $ are shown in red.}
			\label{Breathers_contW}
		\end{figure}

\begin{figure}[h]
			\centering
			\includegraphics[width=\columnwidth]{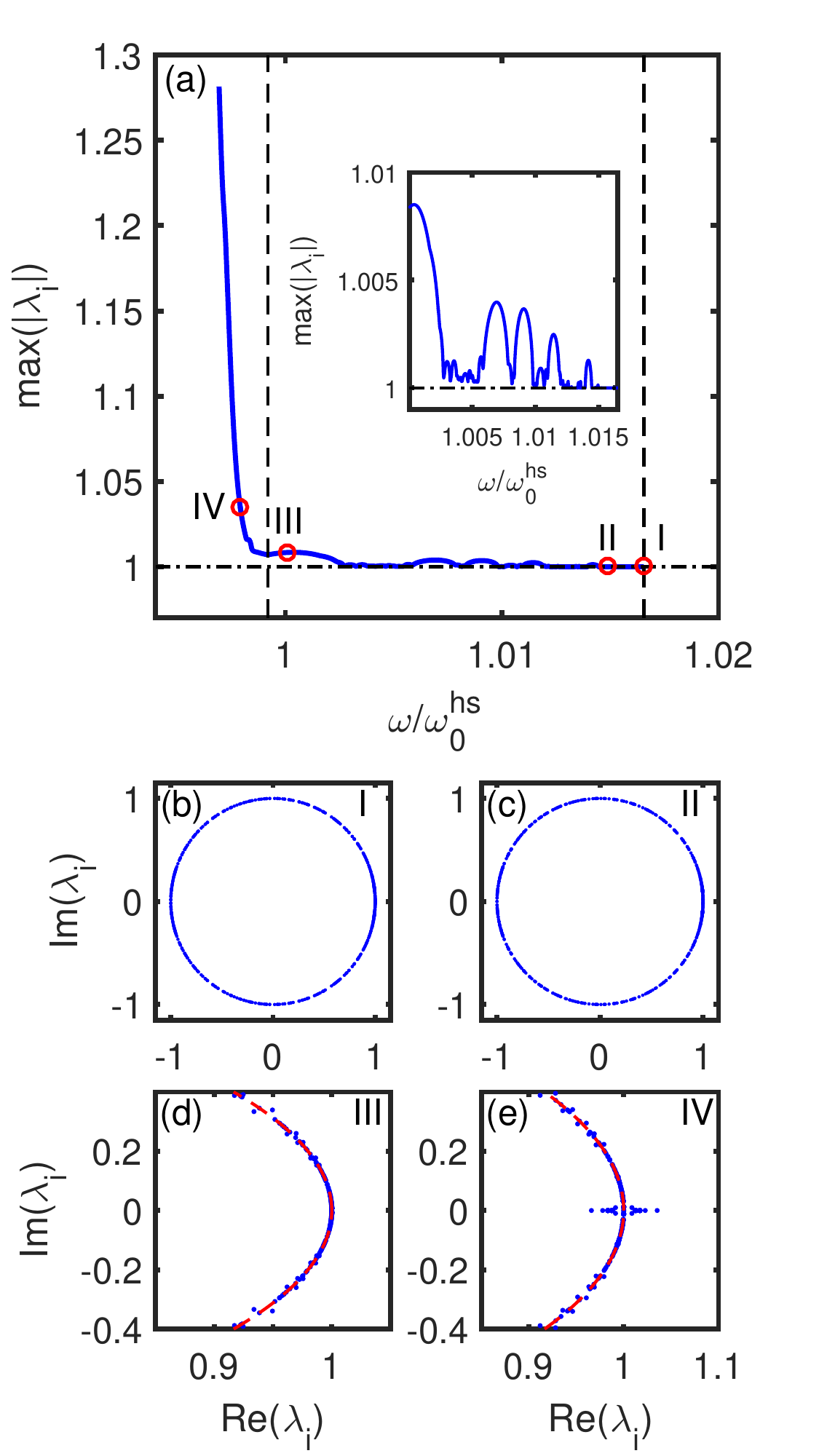}
			\caption{(a) The blue solid curve shows the maximum magnitude of the breather Floquet multipliers. Black dashed lines indicate the edges of the linear phonon band gap, and the inset shows a magnified view of the data in this range.   (b-e) Floquet multipliers (blue dots) of the solutions labeled $ I $ - $ IV $, respectively, in (a), corresponding to the same points in Fig. \ref{Breathers_contW}.  In (d) and (e), axes limits are chosen to emphasize the deviation from the unit circle, which is shown as a visual aid (red dashed lines).}
			\label{Floquet_contW}
		\end{figure}

		As shown in Fig.~\ref{Breathers_contW}, the continuation reveals a family of DBs that extends from the linear eigenmode at vanishing amplitude and traverses the band gap (and into the passband). In Fig.~\ref{Floquet_contW}(a), we show the maximum magnitudes of the Floquet multipliers of the branch.
		 This family of DBs exhibits behavior similar to those found in previous studies of diatomic granular chains \cite{Theocharis2010PRE}, where the DBs are linearly stable for frequencies very close to the lower cutoff frequency of the optical branch. As the frequency decreases, the breather profiles become progressively more localized in space, as can be seen in Fig.~\ref{Breathers_contW}(b-d). The Floquet multipliers corresponding to the modes shown in Fig.~\ref{Breathers_contW}(b-d) are shown in Fig.~\ref{Floquet_contW}(b-d), with small deviations from the unit circle emphasized in the latter panel. Below the band gap, interactions with the acoustic band generate oscillating tails, as is shown in Fig.~\ref{Breathers_contW}(e), and Floquet multipliers depart from the unit circle along the real axis, as shown in Fig. ~\ref{Floquet_contW}(e). We note that boundary effects are significant in the presence of oscillating tails, so the finite-length DBs do not accurately approximate the case of an infinite lattice.
As an aside, we also point out that contrary, e.g., to what is
the case in the work of~\cite{Theocharis2010PRE,Boechler2010PRL} for
a granular crystal, here the dependence of the energy (Hamiltonian)
on the frequency is monotonic, hence in accordance with the
recent criterion of~\cite{extrapgk}, no instability arises 
from changes of monotonicity in this dependence.

\section{Numerical Simulations}

	\begin{figure}[h]
		\centering
		\includegraphics[width=\columnwidth]{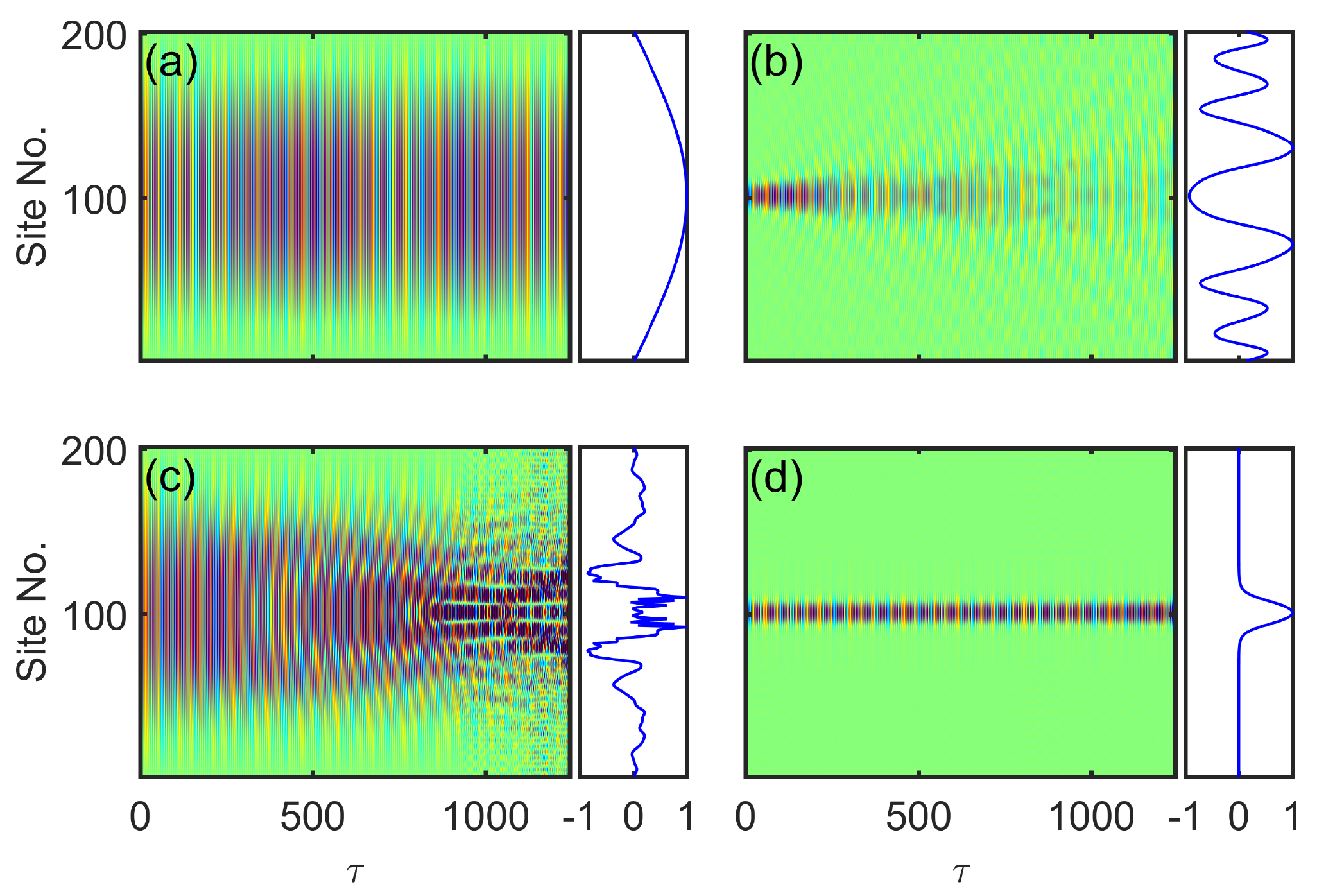}
		\caption{Spatiotemporal plots of the relative displacements $ v_j - u_j $ of the simulated lattice for high and low amplitude excitations, using eigenmode and DB profiles as initial shapes. Side panels contain spatial profiles of $ v_j - u_j $ at the final time step, normalized to the maximum value. (a) Eigenmode shape with low amplitude (approximate periodic solution). (b) DB shape rescaled to low amplitude. (c) Eigenmode shape rescaled to high amplitude. (d) DB shape with high amplitude (exact periodic solution).}
		\label{Fig_Hist}
	\end{figure}

	To explore the dynamics of DBs in our model, we simulate a lattice with 201 unit cells via direct numerical integration of the equations of motion given by Eqs. (\ref{eq: EOM1}) and (\ref{eq: EOM2}).  
	
	
	We consider initial conditions in two shapes: the profile of the DB with frequency $ \omega/\omega_0^{hs}  = 1.01$ and maximum Floquet multiplier magnitude $ \max(|\lambda_i|) = 1.001 $ (as is shown in Fig.~\ref{NLSbreather}(b) and denoted by the star in Fig. \ref{Breathers_contW}(a)), as well as the profile of the seeding eigenmode used in Sec. \ref{SectionCont}. B.
	
	For each of these shapes, we scale the amplitude in two ways. In the case of the DB shape, we consider the exact breather shape computed via continuation (high), and then consider a rescaled DB shape, such that the initial displacement $ v_{101} $ of the local resonator at the central lattice site is equal to one-hundredth of that of the exact solution (low). Similarly, we consider the shape of the seeding eigenmode scaled such that the initial displacement $ v_{101} $ of the local resonator at the central lattice site is matched to the low-amplitude, rescaled DB shape (low), and then finally consider a rescaled eigenmode shape, such that the initial displacement $ v_{101} $ of the local resonator at the central lattice site is equal to that of the exact DB solution (low). Thus, there are four sets of initial conditions: the DB shape with high amplitude (Fig. \ref{Fig_Hist}(d)), which results in an exact periodic solution of Eqs. \eqref{eq: EOM1} and \eqref{eq: EOM2}; the eigenmode shape with low amplitude (Fig. \ref{Fig_Hist}(a)), which closely approximates a periodic solution; the DB shape rescaled to low amplitude (Fig. \ref{Fig_Hist}(b)), which is not a true periodic solution; and the eigenmode shape rescaled to high amplitude (Fig. \ref{Fig_Hist}(c)), which also is not a periodic solution. The duration of all simulations is $ 200T $, where $ T = 2 \pi \omega_0^{hs}/\omega $ is the period of the exact DB solution.
%

	Spatiotemporal plots of the relative displacements $ v_j - u_j $ of the simulated lattice, using the low and high amplitude DB profiles as initial conditions (i.e. the rescaled DB and exact solution) are shown in Fig. \ref{Fig_Hist}(b) and \ref{Fig_Hist}(d), respectively, and the corresponding cases using the eigenmode shape (i.e. the approximate periodic solution and corresponding rescaled profile) are shown in Fig. \ref{Fig_Hist}(a) and Fig. \ref{Fig_Hist}(c). 
	
	%
	As shown in the right column of Fig.~ \ref{Fig_Hist}, the breather shape spreads out from the central lattice sites at low amplitude, but remains highly localized when initiated with the energy of the exact solution. Conversely, as shown in the left column of Fig.~ \ref{Fig_Hist}, the eigenmode shape shows no noticeable distortion at low amplitude, but self-localizes and eventually breaks up at high amplitude. 
In this break-up, many smaller DBs are formed, a process arguably reminiscent of multiple filamentation in nonlinear optics~\cite{couairon}, which also move in space. Thus, in future experiments on this system (e.g. using photoacoustic techniques, as in \cite{Boechler2013PRL}), DBs could be detected by impulsively exciting a large spot on the substrate surface, and observing the formation of smaller, highly localized wave packets. However, we note that we expect that losses, such as may be caused by dissipation and radiation into the substrate, may play an important role in this case, and should be considered in future models.

\section{Conclusion}

In this work, we have demonstrated the existence of discrete breathers in a mass-in-mass chain that models a locally resonant, microscale granular metamaterial composed of microspheres adhered to a substrate. This chain consists of a linearly-coupled main chain (representing the substrate) with nonlinear local resonators that follow the Hertzian contact law (representing the microspheres). After fitting the two independent model parameters to a system studied in previous works, we analyze the resulting energy trapping, in the form of
discrete breathers, theoretically, in the anti-continuum and continuum limits of intersite coupling, as well as numerically, by using the intersite coupling stiffness and oscillation frequency as continuation parameters. 
Finally, we simulate the formation and filamentation--in the form
of discrete breathers--of single-humped wavepackets 
using direct numerical integration of the equations of motion. The simulations suggest that discrete breathers may be observed in experiments on microscale granular metamaterials by generating a long-wavelength SAW at high amplitude, and detecting its breakup into many small DBs.
We expect this work to aid in future studies of nonlinear, microscale granular systems, as well as to the more general class of media composed of a linear material with local nonlinear resonant attachments. Indeed, the dynamics and interactions of discrete breathers (as well as their potential filamentation and dispersion) in one- to three-dimensional analogs of such systems, may yield numerous topics that could be both theoretically intriguing, as well as experimentally accessible for further study. Also, it should be borne in mind that here we only
explored the focusing variants of the relevant models and their bright
solitons. Yet, in line with recent explorations in granular crystals~\cite{jk1}
and in systems with resonators~\cite{Liu2015},
the self-defocusing case may also be interesting and 
equally accessible in different parametric regimes.

N.B. and S.P.W. acknowledge support from the National Science Foundation (grant no. CMMI-1333858) and the Army Research Office (grant no. W911NF-15-1-0030). P.G.K. acknowledges support from the Army Research Office (grant no. W911NF-15-1-0604) and the Air Force Office of Scientific Research (Grant No. FA9550-12-10332).

\appendix
	
	\section{Estimation of the Mass Ratio, $ M $}
	
		The local resonator mass is taken to be the mass of a single microsphere, given by $ m_2 = 4/3 \pi (D/2)^3 \rho_2 $, where $ \rho_2 $ is the density of the microsphere material. We estimate the main chain mass as that of a rectangular region of the substrate beneath a single microsphere. Since Rayleigh SAWs have a characteristic decay length on the order of the wavelength $ \lambda $ \cite{Ewing}, this region has mass $ m_1 =  \lambda D^2 \rho_1$, where $ \rho_1 $ is the density of the substrate material.
		
		The mass ratio is then given by
			 \begin{equation}
				 M = m_1/m_2 = \left(\frac{6}{\pi} \frac{\rho_1}{\rho_2} \right) \frac{\lambda}{D},
			 \end{equation} 
		where the quantity in parentheses is expected to be of order $ \sim 1 $ for most material combinations, and the quantity $ \lambda/D $ is at least of order $ \sim10^1 $, since we consider long wavelengths. Therefore, we estimate $ M \sim 10^1 $.
		\newline
	\section{Estimation of the Stiffness Ratio, $ K $}
		
		The local resonator stiffness is taken from Hertzian contact theory \cite{Hertz}, linearized about the static overlap distance $ \delta_0 $ due to adhesion, using the same model as Ref. \cite{Boechler2013PRL}.  This stiffness is given by $ k_2 = (3/2) A \sqrt{\delta_0} $. Here, $ A = (4/3) E^* \sqrt{D/2} $, with $ E^* = \left[ (1-\nu_1^2)/E_1 + (1-\nu_2^2)/E_2 \right]^{-1} $, where $ E_{1,2} $ and $ \nu_{1,2} $ are the Young's modulus and Poisson's ratio, respectively, with subscripts corresponding to the substrate and sphere materials  \cite{Hertz}. The static overlap is $ \delta_0 = (F_{DMT}/A)^{2/3} $, where $ F_{DMT} =  \pi w D $ is the force due to adhesion as per the Derjaguin-Muller-Toporov (DMT) adhesive elastic contact model \cite{DMT}, and $ w $ is the work of adhesion between the sphere and substrate materials. The main chain stiffness is estimated in a similar manner to the mass $ m_1 $, using an element of the substrate with length $ D $ in the direction of SAW propagation, and cross-sectional area $ \lambda D $.  The estimated stiffness is then $ k_1 = E_1 \lambda $. 
		
		After algebraic manipulation, the stiffness ratio can be written as 
		\begin{equation}
			K = k_1/k_2 \approx \left( \frac{8(1-\nu_1^2)^2}{3 \pi} \right)^{1/3} \left( \frac{E_1 D}{2 w} \right)^{1/3} \frac{\lambda}{D}. 
		\end{equation}
		\noindent
		For simplicity, we have chosen to use identical sphere and substrate materials (so that $ E_2 = E_1 $ and $ \nu_2 = \nu_1 $); this approximation does not affect the generality of this rough estimate. The first term in parentheses is of order $ \sim 10^{-1} $ for realistic values of Poisson's ratio, i.e. $ 0 \leq \nu_1 \leq 0.5 $, and the second term in parentheses is of order $ \sim 10^2 $, for realistic values $ E_1 \sim 10^{10} $ and $ w \sim 10^{-2} $, with $ D \sim 10^{-6} $, using SI units. Since quantity $ \lambda/D \sim 10^1 $,  we estimate $ K \sim 10^2 $.

	
	\section{Details of NLS derivation} \label{NLS_append}
	Define the residuals as
	\begin{eqnarray}
		res(u) = -  M\pa_{tt}u &\nonumber\\
		 + \pa_{xx}u - (u-v) - \frac{1}{4}(u-v)^2+\frac{1}{24}(u-v)^3  &\\ 
		 res(v) =  -\pa_{tt}v  \nonumber&\\ 
		 + (u-v) + \frac{1}{4}(u-v)^2-\frac{1}{24}(u-v)^3.&
	\end{eqnarray}
Substituting $ u^{an}$ and $v^{an} $ of Eqs.~\eqref{longanz1} and \eqref{longanz2}, we obtain the following residuals organized by orders of $\eps$:

\begin{equation} res(u^{an}) = \qquad \qquad \qquad \qquad \end{equation} 
$\varepsilon^{\ \,}[(\omega^2M - k^2 -(1-a_1)   )A\ E] + $\\
$\varepsilon^2[ \left( (4\omega^2M -4k^2 - 1)a_2  +  a_3 - \frac{1}{4}(1-a_1)^2 \right)A^2\ E^2 +$

$\left(2ic\omega M + 2ik +  a_7 \right) \pa_XA E  +\left(  a_{12}- \frac{1}{2}(1-a_1 )^2\right) A\ \bar{A}  
 ]  $ \\
$\varepsilon^3 [(1-Mc^2+a_{11}) \pa_X^2A\ E  + ( (\omega^2M-k^2-1)a_{10} -2i\omega M ) $
\begin{flushright}\vspace{-6pt}$\pa_TA E + $ \end{flushright}

\vspace{-6pt}$ \left(\frac{1}{8}(1-a_1)^3-\frac{1}{2}(1-a_1)(a_2 -a_3-a_{12}))\right)|A|^2A\ E  + $

$(8ic\omega a_2 M+ 8ika_2 +(4\omega^2M-4k^2-1)a_8 +  a_9 +$ 
\begin{flushright} \vspace{-6pt}{$ \frac{1}{2}(1-a_1)a_7) A\pa_XA\ E^2 + $} \end{flushright}

\vspace{-6pt}$((9\omega^2M-9k^2 -1)a_4+  a_5 -\frac{1}{2}(1-a_1)(a_2 -a_3)+$ 
\begin{flushright} \vspace{-6pt}$ \frac{1}{24}(1-a_1)^3)A^3\ E^3   + $ \end{flushright}

\vspace{-6pt}$\left(\frac{1}{2}(1-a_1)a_7 - a_6\right) \bar{A}\pa_XA  ] + c.c. + \mathrm{h.o.t.} $ 

\begin{equation} res(v^{an}) = \qquad\qquad\qquad\qquad \end{equation}
$\varepsilon^{\ \,}[\left(\omega^2  a_1+  (1-a_1) \right) A E]  + $
$\varepsilon^2 [\left(4\omega^2 a_3+  (a_2 -a_3)  + \frac{1}{4}(1-a_1)^2 \right)A^2\ E^2  +$ 

$\left(2ic\omega a_1+ \omega^2 a_7 -  a_7 \right)\pa_XA\ E + $ 

$\left(\frac{1}{2}(1-a_1)^2 - a_{12} \right) A\bar{A} ] + $ 
$ \varepsilon^3[ \left(2ic\omega a_7 - c^2a_1 + \omega^2a_{11} - a_{11} \right)\pa_X^2A E+ $

$\left(a_{10}-2i\omega a_1\right) \pa_TA\ E+$

$\left(\frac{1}{2}(1-a_1)(a_2-a_3-a_{12})-\frac{1}{8}(1-a_1)^3\right)
|A|^2A\ E  +$

$\left( 9\omega^2 a_5+ (a_4-a_5) + \frac{1}{2}(1-a_1)(a_2 -a_3)-\frac{1}{24}(1-a_1)^3\right)$
\begin{flushright}\vspace{-6pt} $A^3E^3 +$\end{flushright}

\vspace{-6pt} 
$ \left(8ic \omega a_3+ 4\omega^2 a_9 + (a_8-a_9) -  \frac{1}{2}(1-a_1)a_7 \right)$
\begin{flushright}\vspace{-6pt} $A\pa_XA\ E^2+$\end{flushright}

\vspace{-6pt} 
$\left( a_6 - \frac{1}{2}(1-a_1)a_7\right) \bar{A}\pa_XA ] + c.c. + \textrm{h.o.t}$  \vspace{6pt}  \\
If we set each order of $\eps$ to $0$, we can define the coefficients, $a_i$ and parameters, $\omega$ and $c$, such that  $res(u),resu(v) = \mathcal(\eps^4)$,
which should yield an accurate approximate solution to our original equations of motion. We now list the heiarchy of solvability conditions:

\noindent $\mathcal{O}(\eps AE):$ the dispersion relation, 
$$\quad M\omega^4 - \left[1+k^2+M\right]\omega^2 + k^2 =0 \quad $$ 
$$\quad a_1 = 1+k^2-M\omega^2$$
$\mathcal{O}(\eps^2 A^2E^2)$:

$$\quad a_2 = -\frac{\omega^2(1-a_1)^2}{1+ (1-4\omega^2)(4 \omega^2M-4k^2-1)}\quad $$
$$\quad a_3 = \frac{1}{4} (1-a_1)^2 -(4\omega^2M-4k^2-1)a_2$$

$\mathcal{O}(\eps^2 \pa_X A \ E):$

$$\quad a_7 = \frac{-2ika_1}{M(1-w^2) +a_1} $$
$$\quad c = -\frac{2ik+a_7}{2i\omega a_1}$$

$\mathcal{O}(\eps^2 A \bar{A}):$

$$\quad a_{12} = \frac{1}{2}(1-a_1)^2$$
 
$\mathcal{O}(\eps^3 A^3E^3)$:

$$\quad a_4 = -\frac{9\omega^2\left(\frac{1}{2} (1-a_1)(a_2-a_3) +\frac{1}{24}(1-a_1)^3\right)}{1-(9\omega^2-1)(9\omega^2M-9k^2-1)} \quad $$ 
$$\quad a_5 = \frac{a_4 +\frac{1}{2}(1-a_1)(a_2-a_3)  -\frac{1}{24} (1-a_1)^3 }{(1-9\omega^2)a_5}$$

$\mathcal{O}(\eps^3 A\pa_xA\ E^2):$
  
$$ \quad a_8 = \frac{8ia_2(k+c\omega M)(4\omega^2-1) +2a_7(1-a_1)\omega^2-8ic\omega a_3 }{1+(4\omega^2M-4k^2-1)(1-4\omega^2)} \quad $$
$$ \quad a_9=\frac{8ic\omega  a_3 +a_8-\frac{1}{2}(1-a_1)a_7}{1-4\omega^2}$$

$O(\eps^3 \bar A \pa_xA):$
$$\quad a_6 = \frac{1}{2} a_7(1-a_1)$$ \\

Finally, with these coefficients, we are left with two NLS equations at $\mathcal{O}(\eps^3 E)$ of both $res(u)$ and $res(v)$.\\
In $res(u)$, we have:
\begin{eqnarray} \nonumber 
(-2i\omega M +(\omega^2M-k^2-1)a_{10})\pa_TA &=\\ \nonumber-(1-Mc^2+a_{11}) \pa_X^2 A &+ \\
\left(\frac{1-a_1}{2}(a_2-a_3-a_{12}) -\frac{(1- a_1)^3}{8}\right)|A|^2& A 
\end{eqnarray}
and in $res(v)$, we have:
\begin{eqnarray}
\nonumber  -(-2i\omega a_1 +a_{10})\pa_TA &=\\
\nonumber (2ic\omega a_7+\omega^2a_{11}-a_{11}-c^2a_1) \pa_X^2 A &+ \\
\left(\frac{(1- a_1)^3}{8}-\frac{1-a_1}{2}(a_2-a_3-a_{12})\right)|A|^2& A
\end{eqnarray}
In order for both NLS equations to be satisfied, the coefficients in each must match. Thus, we require that 
\[a_{10} = \frac{2i\omega(a_1+\omega M)}{\omega^2 M -k^2}  \textrm{  and  }   a_{11}=\frac{Mc^2 -1-2ic\omega a_7 +c^2a_1}{w^2} \]
so that we obtain a single NLS equation. This equation has the solution 
\[A(X,T)=\sqrt{\gamma} \alpha \sech(\sqrt{\gamma} \beta X)\ee^{-i\gamma T} \]
with \begin{eqnarray}
\gamma \beta^2      &=&  \frac{2\omega M+(\omega^2M-k^2-1)a_{10}i   }{1-Mc^2+a_{11}}  \\
\gamma \alpha^2      &=&  \frac{4\omega M +2(\omega^2M-k^2-1)a_{10} i} {\frac{1}{8}(1-a_1)^3-\frac{1}{2}(1-a_1)(a_2 -a_3-a_{12}) }
\end{eqnarray}
and $\gamma$ a free parameter.

\end{document}